\documentstyle[12pt]{article}
\input{tcilatex}

\begin{document}

\vspace{-1cm}

\title{\large\bfseries{PHYSICAL INTERPRETATION OF THE MATHEMATICAL \\
CONSEQUENCE OF LORENTZ' TRANSFORMATIONS}}

\author{Josiph Mladenov Rangelov, \\
Institute of Solid State Physics , Bulgarian Academy of \\
Sciences , blv.Tzarigradsko Chaussee 72 ,Sofia 1784 , Bulgaria}

\date{}

\maketitle

\begin{abstract}
  A physical interpretation of the mathematical consequence of Lorentz
transformation within spatial relativity theory is presented as a result
of my new physical model of existent fluctuating vacuum (FlcVcm). It is
assumed that the FlcVcm is considered as molecular dielectric, which
consists from neutral dynamides, streamlined in a close-packed crystalline
lattice.  Every dynamide is a neutral pair, consistent by two massless
opposite point-like elementary electric charges (ElmElcChrgs): electrino (-)
and positrino (+). In a frozen equilibrium position two contrary pont-like
ElmElcChrgs within every one dynamide are very closely installed one to
another and therefore the aggregate polarization of every dynamide and its
electric field also have zero values. The aggregate electric field of every
dynamide polarizes nearest neighbors dynamides in an account of which nearest
dynamides interact between them-self, because of which their elementary
excitations, phonons and photons, have a wave character and behaviors. We
suppose that the photon is an polarization result of the phonon within the
fluctuating vacuum considered as an ideal dielectric and therefore the
photon could be considered as an elementary collective excitation of the
FlcVcm in the form of a solitary needle cylindrical harmonic oscillation.
Hence the light, which is a packet of the photons, must move within FlcVcm
with constant velocity and Dopler effect must be observed in both cases,
for the light and sound. Then all mathematical results of Lorentz
transformation could be considered as results of a demand of an independence
of the observation results from the reactive velocity of the observation frame.
\end{abstract}

  A physical model (PhsMdl) of the existent fluctuating vacuum (FlcVcm) and
its elementary excitations phonon and photon as a solitary needle cylindrical
harmonic oscillation of its polarized molecular crystal lattice is offered.
It is common known that the physical model (PhsMdl) presents at us as an
actual ingredient of every good physical theory (PhsThr). It would be used as
for an obvious visual teaching the unknown occurred physical processes within
the investigated phenomena. We assume that the FlcVcm could be considered as
a molecular crystal, which is consistent by neutral dynamides, streamlined in
some close-packed crystalline lattice. Every dynamide is a massless neutral
pair, consistent by two massless opposite point-like (PntLk) elementary
electric charges (ElmElcChrgs): electrino (-) and positrino (+). In a frozen
equilibrium position both opposite PntLk ElmElcChrgs within every dynamide
are very closely installed one to another and therefore the aggregate
polarization of every one dynamide has zero value and its electric field
(ElcFld) also has zero electric intensity (ElcInt). However the absence of a
mass in a rest of the electrino and positrino makes them possible to have a
big mobility and infinitesimal dynamical inertness of its own QntElcMgnFld,
what permits them to be found a bigger time in an unequillibrium distorted
position. The aggregate ElcFld of the dynamide reminds us that it could be
considered as the QntElcFld of an electric quasi-dipole moment (ElcQusDplMmn)
because both opportunity massless electrino and positrino have the same
inertness. For a certain that is why the FlcVcm dos not radiate real photon
(RlPhtn) by itself, as a dynamide electric dipole moment (ElcDplMmn) has a
zero value. The aggregate ElcFld of every dynamide polarizes nearest
neighbors dynamides in an account of which nearest dynamides interact between
itself, and in a result of which its elementary collective excitations have a
wave character and behavior. It is richly clear that the motions in the
opposite direction of both opposite PntLk ElmElcChrgs of an every dynamide
creates an aggregate magnetic field (MgnFld) of every one and the sum of
which makes a magnetic part of the free QntElcMgnFld.

  Although up to the present nobody of scientists distinctly knows are there
some elementary micro particles (ElmMicrPrts) as a fundamental building
stone of the micro world and what the elementary micro particle (ElmMicrPrt)
means, there exists an essential possibility for physical clear and
scientific obvious consideration of the uncommon quantum behavior and
unusual dynamical relativistic parameters of all relativistic quantized
MicrPrts (QntMicrPrts) by means of our convincing and transparent surveyed
PhsMdl. We suppose that the photon is some elementary excitation of the
FlcVcm in the form of a solitary needle cylindrical harmonic oscillation.
The deviations of both PntLk massless opportunity ElmElcChrgs of an every
dynamide from their equilibrium position in the vacuum close-packed
crystalline lattice creates its own polarization, the sum of which creates
total polarization of the FlcVcm as a ideal dielectric, which causes the
existence of a total resultant QntElcFld. Consequently the total
polarization of all dynamides creates own resultant QntElcFld, which is an
electric part of the free QntElcMgnFld. Really, if the deviation of an every
PntLk ElmElcChrg within every one dynamide from its own equilibrium position
is described by dint of formula of collective oscillations (RlPhtns) of
connected oscillators in a representation of second quantization:
\begin{equation}  \label{a}
u_{j}(r)\,=\,\frac{1}{\sqrt{N}}\sum_{q}\sqrt{\frac{\hbar }{2\Theta \omega }}
\,I_{jq}\,\left\{ \,a_{jq}^{+}\exp {\,i(\,\omega t\,-\,qr\,)}
\,+\,a_{jq}\,\exp {-i(\,\omega t\,-\,qr\,)}\,\right\}
\end{equation}

where $\Theta$ is an inertial mass of the electrino and positrino and $%
I_{jq} $ are vector components of the deviation (polarization). If we
multiply the deviation u of every PntLk ElmElcChrg in every dynamide by the
twofold ElmElcChrg value e and dynamide density $W=\frac{1}{\Omega _{o}}$,
then we could obtain in a result the total polarization value of the FlcVcm
within a representation of the second quantization :
\begin{equation}  \label{b}
P_{j}(r)\,=\,\frac{2e}{\Omega_o \sqrt{N}} \sum_{q} \sqrt{\frac{\hbar}{%
2\Theta \omega}}\,I_{jq}\,\left\{\,a_{jq}^{+}\,\exp{\,i\,(\omega t\,-\,qr)}%
\,+\, a_{jq}\,\exp {-\,i\,(\omega t\,-qr\,)}\,\right\}
\end{equation}

  Further we must note that the change of the spring with an elasticity $\chi$
between the MicrPrt and its equilibrium position, oscillating with a
circular frequency $\omega $ by two springs with an elasticity $\tilde{\chi}$
between two MicrPrts, having opportunity ElmElcChrgs and oscillating with a
circular frequency $\tilde{\omega}$ within one dynamide, is accompanied by a
relation $\,2\,\tilde{\chi}\,\simeq\,\chi$. Indeed, if the ,,masses'' of the
oscillating as unharmed dynamide is twice the ,,mass'' of the electrino or
positrino, but the elasticity of the spring between every two neighbor
dynamides in crystalline lattice is fourfold more the elasticity of the
spring between two the MicrPrts, having opportunity ElmElcChrgs and
oscillating one relatively other within one dynamide, while the common
,,mass'' of two the MicrPrts, having opportunity ElmElcChrgs and oscillating
one relatively other within one dynamide is half of the ,,mass'' of the
electrino or positrino. Therefore the circular frequency $\omega $ of the
collective oscillations have well known relation with the Qoulomb potential
of the electric interaction (ElcInt) between two opportunity massless PntLk
ElmElcChrgs electrino and positrino and their dynamical inertial ,,masses''
which can be described by dint of the equations :
\begin{equation}  \label{c1}
\tilde{\omega}^{2}=2\frac{\tilde{\chi}}{\Theta}\quad{\rm and}\quad
\omega^{2}\,= \,\frac{4\chi }{2\Theta }\quad {\rm consequently}\quad \omega
^{2}\,= \,2 \tilde{\omega}^{2}
\end{equation}

\begin{equation}  \label{c2}
\quad {\rm and}\quad {\rm therefore}\quad \Theta \omega ^{2}=\frac{4e^{2}} {%
4\pi \Omega _{o}\,\varepsilon _{o}}\quad {\rm or}\quad \Theta C^{2}\,= \,%
\frac{e^{2}}{4\pi \Omega _{o}\,q^{2}\,\varepsilon _{o}}
\end{equation}

where
\begin{equation}  \label{d}
\,N\Omega _o\,=\,\Omega \quad {\rm and} \quad d = W\,e\,E \quad {\rm or}
\quad \,E = \frac{d}{\Omega _o\varepsilon _o} = \frac{P}{\varepsilon _o}
\end{equation}

we could obtain an expression for the ElcInt of the QntElcMgnFld, well known
from classical electrodynamics (ClsElcDnm) in a representation of the second
quantization:
\begin{equation}  \label{e}
E_{j}(r)=\sum_{q}\sqrt{\frac{2\pi\hbar\omega}{\Omega\varepsilon_{o}}}\,
I_{jq}\,\left\{ \,a_{jq}^{+}\,\exp {i(\omega t-\,qr)}\; +\;a_{jq}\,\exp {%
-i(\omega t-\,qr)}\right\}
\end{equation}

  By dint of a common known defining equality :
\begin{equation}  \label{f}
E_j = -\,\frac{\partial A_j}{\partial t}
\end{equation}

  From (\ref{f}) we could obtain the expression for the vector-potential A of
the QntElcMgnFld in the vacuum in a representation of the second
quantization :
\begin{equation}  \label{g1}
A_{j}(r)=i\,\sum_{q}\,\sqrt{\frac{2\pi\hbar}{\Omega\omega\varepsilon_{o}}}\,
I_{jq}\,\left\{ \,a_{jq}^{+}\,\exp {\,i(\,\omega t-qr\,)}-\,a_{jq}\, \exp{%
-\,i(\,\omega t-qr)}\right\} \,
\end{equation}

or
\begin{equation}  \label{g2}
A_{j}(r)=i\,\sum_{q}\,\sqrt{\frac{2\pi\hbar\omega\mu_{o}}{\Omega q^{2}}}\,
I_{jq}\,\left\{ \,a_{jq}^{+}\,\exp {\,i\,(\omega t-qr\,)}\,-\,a_{jq}\, \exp{%
-\,i\,(\omega t-qr)}\right\}
\end{equation}

  Further by dint of the defining equality $\mu _{o}H=rotA$ from (\ref{g1})
and (\ref{g2}) we could obtain an expression for the MgnInt of QntElcMgnFld,
well known from ClsElcDnm in a representation of the second quantization:
\begin{equation}  \label{h}
H_{j}(r)=\sum_{q}\,\sqrt{\frac{2\pi \hbar \omega }{\Omega \mu _{o}}}\, \left[%
\,\vec{n}_{q}\times I_{lq}\,\right] _{j}\left\{ \,a_{jq}^{+}\, \exp{%
\,i(\,\omega t-qr)}+a_{jq}\,\exp {-\,i(\,\omega t-qr)}\right\}
\end{equation}

where $\vec n_k$ is unit vector, determining the motion direction of the
free QntElcMgnFld. The reception of known expressions for the ElcInt and
MgnInt values of the QntElcMgnFld by dint of a simple transformation of an
expression, describing deviation of two PntLk ElmElcChrgs of distorted
dynamides into the ideal dielectric of the FlcVcm proves obviously and
scientifically the true of our assumption about the dipole structure of the
vacuum and about the creation way of its collective oscillation - RlPhtn.
The existence of a possibility for a creation of virtual photons (VrtPhtns)
as an polarized excitation within the fluctuating vacuum (FlcVcm) renders an
essential influence over the motion of a electric charged or magnetized micro
particles (MicrPrts) by means of its EntElcMgnFld. The existence of a free
energy in the form of micro particles (MicrPrts) can break of the connection
between pair contrary PntLk ElmElcChrgs of one dynamide and to excite pair of
two opposite charged MicrPrts at once.

  By means of presentation (\ref{h}) of MgnInt $H_j(r)$ and taking into
consideration that $\vec n_q$ is always perpendicular to the vector of the
polarization $I_{jq}$ we obtain that :
\begin{equation}  \label{ha}
\,\left[\,\vec v\,\times \left[\,\vec n_q \times \,I_{jq}\right]\,\right] =
\,\vec n_q\,(\vec v\,\cdot\,I_{jq}\,) - I_{jq}\,(\,\vec v\,\cdot\,\vec n_q\,)
\end{equation}

  From this equation (\ref{ha}) we could understand that if the velocity $v$
of the interacting ElcChrg is parallel of the direction $\vec{n}_{q}$ of the
motion of the free QntElcMgnFld, then the first term in the equation (\ref
{ha}) will been nullified and the second term in the equation (\ref{ha})
will determine the force, which will act upon this interacting ElcChrg. But
when the velocity $\vec{v}$ of the interacting ElcChrg is parallel of the
direction $I_{jq}$ of the motion in the opposite directions of two PntLk
ElmElcChrg of the electrino and positrino and one is a perpendicular to the
direction $\vec{n_{q}}$ of the motion of a free QntElcMgnFld, then the
second term in the equation (\ref{ha}) will be nullified and the first term
in the equation (\ref{ha}) will describe the force, which acts upon this
interacting PntLk ElmElcChrg. It turns out that the interaction between
currents of the electrino and positrino, which is parallel to the vector of
a polarization $I_{jq}$ as $(\vec{v}_{j}=\frac{\omega }{\pi }\,\vec{I}_{jq})$%
, with the QntMgnFld of the free QntElcMgnFld determines the motion and its
velocity of same this free QntElcMgnFld. Indeed, it is well known that the
change of a magnetic flow $\Phi $ creates a ElcFld. Therefore by dint of a
relation (\ref{ha}) we can obtain the following relation :
\begin{eqnarray}  \label{hb}
F_{j} = \frac{e}{C}\left[ \vec{v}\times \vec{H}\right] =\frac{e}{m\,C\,\omega%
} \left[ \vec{E}\times \vec{H}\right] = \sum_{q}\frac{e^{2}}{m\,C}\, \sqrt{%
\frac{2\,\pi\,\hbar}{\Omega\,\varepsilon_{o}}}\sqrt{\frac{2\,\pi\,\hbar} {%
\Omega\,\mu _{o}}}\epsilon _{jkl}\vec{n}_{j}(\vec{I}_{kq}\cdot\vec{I}_{lq})
\nonumber \\
\left\{\,a_{kq}^{+}\,\exp{i(\omega t - qr)} + a_{kq}\,\exp {-i(\omega t - qr)%
} \,\right\}\left\{\,a_{kq}^{+}\exp{i(\omega t - qr)} - a_{kq}\, \exp{%
-i(\omega t - qr)}\,\right\}
\end{eqnarray}

  Therefore by dint of (\ref{hb}) and defining equations (\ref{e}) and (\ref
{g2}) we can obtain:
\begin{equation}  \label{i1}
\frac{1}{\sqrt{\varepsilon \varepsilon _o}} = v \sqrt{\mu \mu _o}\quad{\rm or%
} \quad {\rm at} \quad \frac{1}{\sqrt{\varepsilon _o}} = C \sqrt{\mu _o}%
\quad {\rm we \quad have} \quad C = v . \sqrt{\varepsilon \mu}.
\end{equation}

  It is naturally that when some RlPhtn is moving within the space of some
substance, then supplementary polarization of atoms and molecules appears ,
which delay its moving and slow down its velocity. Indeed in this case the
dielectric constant $\varepsilon$ has a following form:
\begin{equation}  \label{i2}
\varepsilon = 1 + \sum_q \frac{4 \pi n(q)[\omega(q) - \omega_c]}{\left[\,m\,
\{4(\omega(q) - \omega_c)^2 + \tau^2\,\omega(q)^4\}\right]}
\end{equation}

  By means of the upper scientific investigation we understand that the
creation of the QntMgnFld by moving oposete PntLk ElmElcChrgs of electrinos
and positrinos within all dynamides together with their aggregate QntElcFld
as two components of one free QntElcMgnFld one secures their motion.
Therefore we should write the momentum of the free QntElcMgnFld by means of
the equation of Pointing/Umov, using the definition equations (\ref{e}) and (%
\ref{h}) :
\begin{eqnarray}  \label{j1}
&P = \frac{\left[ E\times H\right]}{4\pi C^{2}} = \sum_{q}\vec{n}_{q}
\frac{\displaystyle{\hbar \omega}}{\displaystyle{2\Omega \sqrt{\varepsilon_{o}
\mu_{o}}}}(I_{jq}\cdot I_{jq})&  \nonumber \\
&\left\{a_{jq}^{+}\exp{i(\omega t - qr)}\,+\,a_{jq}\,\exp{-i(\omega t - qr)}
\,\right\}\times &  \nonumber \\
&\left\{\,a_{jq}^{+}\exp{i(\omega t - qr)}\,+\,a_{jq}\, \exp{-i(\omega t -
qr)}\right\} &
\end{eqnarray}

or
\begin{equation}  \label{j2}
P = \sum_q \vec{n}_q \frac{\hbar\,\omega}{2\,\Omega\,C} \left\{\,a_{jq}^{+}
a_{jq} + a_{jq} a_{jq}^{+} + a_{jq}^{+} a_{jq}^{+} \exp{2i(\omega t - qr)} +
a_{jq} a_{jq} \exp{-2i(\omega t - qr)}\,\right\}
\end{equation}

or
\begin{equation}  \label{j3}
\bar P = \sum_q \vec{n}_q \frac{\hbar \omega }{\Omega\,C} (n_q + 1/2)
\end{equation}

  It is well known that the ElmMicrPrts behavior would be studied by means of
an investigation of their behaviors after their interaction by already well
known ElmMicrPrts. Therefore we shall describe the properties and behavior
of the real photon (RlPhtn) by means of a new physical interpretation of
results of its emission and absorption from atoms at their excitatived
Schrodinger electrons (SchEls) from higher energetic state into lower
energetic state or vice versa transition. In such a way we could understand
the origin of some their name by dint of the physical understanding these
determining processes.

  In a first we begin by supposing that the RlPhtn has a form of a solitary
needle cylindrical harmonic soliton with a cross section $\sigma_1$,
determined by the following equation :
\begin{equation}  \label{k1}
\sigma_1 = \pi \{(\delta x)^2 + (\delta y)^2\} = \pi \left[\frac{C}{\omega} %
\right]^2 = \frac{2}{\pi}\left(\frac{\lambda}{2}\right)^2 ,
\end{equation}

which is determined by Heisenberg uncertainty relations:
\begin{equation}  \label{k2}
(\delta p_x)^2 (\delta x)^2 \simeq \frac{\hbar^2}{4} ; \qquad {\rm and}
\qquad (\delta p_y)^2 (\delta y)^2 \simeq \frac{\hbar^2}{4} ;
\end{equation}

where the dispersions are:
\begin{equation}  \label{k3}
(\delta x)^2 \simeq (1/2)\{\frac{C}{\omega}\}^2 = \{\frac{\lambda}{2\pi}%
\}^2; \qquad(\delta y)^2 \simeq (1/2)\{\frac{C}{\omega}\}^2 = \{\frac{\lambda%
}{2\pi}\}^2;
\end{equation}

  It is well known that the probability $P_{12}$ for a transition par second
of some SchEl under ElcIntAct of extern ElcMgnFld from an eigenstate 1 into
an eigenstate 2 is determined by the following formula:
\begin{equation}  \label{k4}
P_{12} = \frac{4}{3} \cdot \frac{e^2}{\hbar C^3}\; ( \omega_{12} )^3\;
|\;\langle\;1\;|\;r\;|\;2\;\rangle\;|^2
\end{equation}

  As the intensity of the ElcMgn emission I, emitted par second is equal of
the product of the probability $P_{12}$ for a transition par second by the
energy $\hbar \omega_{12} $ of the emitted RlPhtn, then for certain
\begin{equation}  \label{k5}
I = P_{12} \hbar \omega_{12} = \frac{4}{3} \frac{e^2}{C^3} {\omega_{12}}^4
\; |\;\langle \;1 \;|\; r \;|\; 2 \;\rangle \;|^2
\end{equation}

  Really the matrix element $\langle 1|r|2 \rangle$ of the SchEl position is
determined by the product of the probability for the spontaneous transition
of a SchEl from an higher energetic level into a lower energetic level and
the number $(n+1)$ for the emission or the number $n$ for the absorption of
a RlPhtn, where n is the number of the RlPhtns within the external
QntElcMgnFld, which polarizes atom. In our view here we need to note obvious
supposition that the spreading quantum trajectory of the SchEl is a result
of the participating of its well spread (WllSpr) ElmElcChrg in isotropic
three dimensional (IstThrDmn) nonrelativistic quantized (NrlQnt) Furthian
stochastic (FrthStch) circular harmonic oscillations motion (CrcHrmOscsMtn),
which is a forced result of the electric interaction (ElcIntAct) of the
SchEl's WllSpr ElmElcChrg by the electric intensity (ElcInt) of the
resultant resonance QntElcMgnFld of all stochastic virtual photons
(StchVrtPhtns), existing in this moment of time within the area, where it is
moving. In order to understand this uncommon stochastic motion we must
remember the IstThrDmn nonrelativistic classical (NrlCls) Brownian
stochastic (BrnStch) trembling harmonic oscillation motion (TrmHrmOscMtn) (
\cite{JMRa}), (\cite{JMRb}), (\cite{JMRc}), (\cite{JMRd}), (\cite{JMRe}), (
\cite{JMRf}).

  It is well known that hundred and fifteen years ago (in 1883) Waldemar
Foght have pointed,that the transition from one space-time variables to new
space-time ones my means of relations :
\begin{equation}  \label{l1}
\quad x\prime = x - v\times t \;
\quad y\prime = y \; \quad z\prime = z \;
\quad t\prime = t - \frac{v . x}{C . C} \;
\end{equation}

preserves the form of the wave equation :
\begin{equation}  \label{l2}
\quad \frac{\partial^2 \varphi}{(\partial x)^2} \quad +
\quad \frac{\partial^2 \varphi}{(\partial y)^2} \quad +
\quad \frac{\partial^2 \varphi}{(\partial z)^2} \quad -
\quad \frac{1}{C^2}\frac{\partial^2 \varphi}{(\partial t)^2} \quad =
\quad 0 \;
 \end{equation}

by means of which the electromagnetic wave of the light is described. But if
J.Fidjerald, J.Larmor and A.Lorentz have considerate these transitions as a
result of some physical cause and mathematical consequence, then A.Poancare
and A.Einstein have voice a supposition for a covariancy of the differential
equation (\ref{l2}) in a relation to transition from one  coordinate frame K
to an other coordinate one $K^\prime$, which is moving in relation to first
by velocity v. It turns out that we are able to give some physical
explanation of uncommon mathematical results of the transition (\ref{l2})
within the spatial relative theory by some obvious physical interpretation of
negative experimental results of Mikelson (1881) and Mikelson and Morly
(1887) experiments. We need to point that the presence of the light velocity
C in equation (\ref{l2}) is a result of description of the electromagnetic
field despreading in the space by time. The applicability of Lorentz'
transformation for a determination of the energy-impulse values within
different coordinate frames proves the electromagnetic self-interaction cause
of the micro particles self-energy. In this light we understand why light
velocity presents in  the wave equation (\ref{l2}) .  We will to explain a
  physical sense of the covariance of the wave equation (\ref{l2}) in a
relation ot the Lorentz' transformation (\ref{l1}). This covariance displays
wave processes are covariant to Lorentz' transformation.  Consequently, both
observers of same wave processes, presenting in different coordinate frames,
must seen and describe one and same wave processes. It is turned that for
this aim we need to suppose some obvious results. In first we need to suppose
that the times have one and same value in different space points but within
unmoving coordinate frame. Then it is turn that the times need to have
different values in different space points within moving coordinate frame.
Indeed, if we observe some harmonic oscillator, then we will see it as a
participating in some wave motion within a moving coordinate frame. Then in
order to describe this wave motion as same  harmonic oscillation motion we
need to introduce time, moving by a fase velocity in a relation to the moving
coordinate frame. In other words the demand of the unity of an observation
pictures of wave motions from both observers, presenting in different
coordinate frames, moving by a constant velocity one to other, demands to
losing the unitary of time velocity in  different space points within the
moving coordinate frame.  It is time to begin by clarification of the
  physical sense of the Lorentz' transformation (\ref{l1}).It is well known
from mechanics that the transition from one coordinate frane to unmoving
another one take place by turning one flatness of a real angle $\varphi$ up
to total coincidence of both axis. But from Lorentz' transformation we seen
that in this case we need to turn one flatness of an imagine angle $\varphi =
i \alpha$ up to total coincidence of both axis (ox - a direction of mutualy
moving, ot - a direction of a time), having value, determinate by following
equality : $\tan(\varphi)$ = $i \tanh(\alpha)$ = $i \gamma \frac{v}{C} $. From
this discussion we understand that if the angle $\varphi$ can have values in
area $( 0 , \frac{\pi}{2})$, then the angle $\alpha$ can have values in area
$(0 , \infty)$. This means that the behaviors of the same wave motion in
both coordinate frame must be described in different space points $(r and
r\prime$ and times values $(t and t\prime)$ as in different coordinate
frames the same event of wave motion have different coordinates $(x,t)$ and
$(x\prime,t\prime)$ within the space-time, which are connected by
Lorentz' transformation (\ref{l1}).

  In what followed we will see one very unexpected consequence from Lorentz'
transformation. Indeed,by means of the transformation (\ref{l1})we could
obtain :
\begin{equation}  \label{l3}
\quad dx\prime = \gamma ( dx - v.dt ) \;   \qquad
dt\prime = \gamma ( dt - \frac{v.dx}{C.C}) \;
\end{equation}

  If we derive the first equation within (\ref{l3}) by second one then we
can obtain a following equation:
\begin{equation}  \label{l4}
\quad u\prime = \frac{( u - v )}{ 1 - \frac{u.v}{C.C}}
\end{equation}

if we use $u\prime = \frac{dx\prime}{dt\prime}$ and $u =
\frac{dx}{dt}$.  When within three dimension space the relative velocity v of
the moving inertial coordinate frame has only one component $ v (v_x, 0 , 0 )
$ then we obtain :
\begin{equation} \label{l5}
\quad u\prime_x = \frac{(u_x - v_x)}{ 1 - \frac{u_x v_x}{C.C}} \;
\quad u\prime_y = \frac{u_y}{ 1 - \frac{u_x v_x}{C.C}} \;
\quad u\prime_z = \frac{u_z}{ 1 - \frac{u_x v_x}{C.C}} .
\end{equation}

  It is very easy to seen from equation (\ref{l5}) that when the velocity v
of the relatively motion reach its maximal values C, then the velocity
$u^(\prime)$ is equal of the light velocity C. This means that the maximal
value of the velocity u within every coordinate frame could be equal of the
light velocity C. This common relation is a nature result of the light
spreading invariancy within vacuum in a relation of observer in different
inertial coordinate frame, which is described by Lorentz' transformation.

  However it is no sole uncommon consequence of Lorentz' transformation. Let
we suppose that the values of both times are equal to zero when both
zentrums $0$ and $0\prime$ of both inertial coordinate frames coincident.
We need to point that we may make the equal time values by synchronization of
clocks in all points of the space within unmoving coordinate system. The
concept of time simultaneous within unmoving coordinate system means that
the time values in all points of its space are equal. However their
corresponding time values within the different space points within moving
coordinate system will be different. This means that the time in moving
coordinate systems moving by a phase velocity $v_f = \frac{C^2}{v}$. We
see that the time in unmoving coordinate frame is moving by infinitely
big velocity, which give us possibility to think that the time is equal in
all point of the space within this coordinate system.

  It is very easy to verify that the time interval, measured by two clocks,
being unmoving in moving and unmoving coordinate frames will be difference.
Let the unmoving clock in unmoving coordinate frame takes pase $x = x_1$,
and the unmoving clock in moving coordinate frame takes other place
$x^(\prime) = (x^(\prime))_1$. Then we obtain following relation by means of
Lorentz' transformation
\begin{equation}  \label{l6}
\Delta \tilde t\prime = \gamma . \Delta t
\end{equation}

  But mean-while the clock within unmoving coordinate frame is measured the
time interval $\Delta t = t_2 - t_1$, the time beginning will pass a distance
$\Delta x = - v.\Delta t$ in moving coordinate frame, measured in unmoving
coordinate frame. Consequently the clock in moving coordinate frame, which
is unmoving in this coordinate frame, will pass from the point $x_1$ to the
point $x_2$, where $x_2 = x_1 (1 - \frac{v.v}{C.C})$. In this way after
substitution we could obtain :
\begin{equation}  \label{l7}
\Delta \bar t\prime = \gamma (1 - \frac{v.v}{C.C}) \Delta t =
\frac{\Delta t}{\gamma}
\end{equation}

  Consequently, if we take into consideration the motion of the clock, being
unmoving in the beginning of  the moving coordinate frame in a relation to
unmoving coordinate frame and the motion of the time by the phase velocity
$u_f$  in the moving coordinate frame, then we could obtain :
\begin{equation}  \label{l8}
\Delta t = \gamma . \Delta t\prime
\end{equation}

   Consequently, the difference in display of both clocks a result of both
motions : of the clock motion by velocity v together with the moving
coordinate frame and the time motion by phase velocity $v_f$ in same one.
Indeed, the first time interval is less  $\gamma$ times and the second time
interval is more $\frac{1}{\gamma}$ times. Therefore their product is equal
of quadrate of own true time: $ (\Delta t)^2 = (\Delta \tilde t\prime).
(\Delta (\bar t\prime)$

  In some way we could explain if we want to understand why both length
$\Delta l$ and $\Delta (\tilde l\prime)$ in moving and unmoving coordinate
frames. It is very clear that for a measure the moving train by velocity v
we need to know where are its ends in equal time values, i.s. at
$(\Delta t\prime)$ . Indeed, we need to take into consideration the time
value $\Delta t = \frac{v.l}{C.C}$ of spread in the time within moving
coordinate frame between both train ends.Therefore after substitution we can
obtain:
\begin{equation}  \label{l9}
\tilde l = \gamma (1 - \frac{v.v}{C.C}) l = \frac{l}{\gamma}
\end{equation}

  If we take into consideration the motion of the beginning of the moving
 coordinate frame toghether by one its end then we could obtain a relation:
\begin{equation}  \label{l10}
(\bar l\prime) = \gamma (1 - \frac{v.v}{C.C}) l = \frac{l}{\gamma}
\end{equation}

 and
\begin{equation}  \label{11}
l = \gamma (\bar l\prime)
\end{equation}

  Therefore the following relation is appeared:
\begin{equation}  \label{l12}
(l\prime)^2 = (\tilde l\prime) (\bar l\prime)
\end{equation}

   The spreading of the proper time within moving coordinate frame finds the
experimental confirmation by the measure of the $\mu$-meson number,
overcoming from the cosmos through the atmosphere to earth surface. Really,
if we take into consideration the live time interval of $\mu$-mesons, within
the unmoving relatively it coordinate system $(\tau = 2,2 . 10^(-6)$ sec,
then although their velocity is smaller then light velocity C, then their
length of half-disintegration will be $l = \tau C = 6,6 10^4 sm = 660 m$.

  Although it is well known that the atmosphere thickness around the earth is
more bigger, a great number of $\mu$-mesons reach it. For a true physical
explanation of this experimental results we need to take into consideration
the time spreading through the space of the moving coordinate frame, which
is unmoving relatively the earth. In such way if we take into consideration
that $\gamma = 100$, then we obtain the value for the time of a $\mu$-meson
half-disintegration $(\tau)^(\prime) = \gamma \tau = 100 2,2 10^(-4) = 2,2
10^(-2)$.  Therefore for the length of the $\mu$-meson half-desintegration we
could obtain  $l = 66,000 m = 66 km $. This example display us that for
understanding the experimental fact in one case we need to increase the train
length and in other cas to increase the half-disintegration time. In such a
physically clear and mathematically correct way we made on progress in
understanding this experimental result.

  It is a time to explain Mikelson-Morly negativ experimental dates.We suggest
that in first an experimental installation is  built in an unmoving inertial
coordinate frame.Moreover the semitransparent mirror (SmMr) is located in the
beginning same inertial coordinate frame and axis ox is directed in parallel
of the velocity v of the moving inertial coordinate frame relatively unmoving
one. Moreover one flat mirror (FlMr1) is installed along axis ox of a
distance l from the beginning of the unmoving inertial coordinate frame and
an other flat mirror (FlMr2) is installed along axis oy, which is a
perpendicular of the axis ox, of a same distance l. The wave front of the
light semi-wave, which is passed through the SmMr at a time moment $t_o =
0$, located in the beginning of the unmoving coordinate frame, arrives in the
FlMr1 and reflects from it at a time moment $t_(1a) = \frac{l}{c}$ and
arrives in the SmMr and reflects from it in direction of axis -oy at a time
moment $t_(2a) = \frac{2l}{C}$. The other wave front of the other light
semi-wave, which reflects from the SmMr in a direction of axis oy and at
same time moment $t_o = 0$, arrives in the FlMr2 and reflects from it at
a time moment $t_(1b) = \frac{l}{C}$ and arrives and passes throgh it at a
time moment $t_(2b) = \frac{2l}{C}$. After all that both light semi-waves
move parallel and arrive to a common screen. It is a classical description.
In more correct quantum description the light wave i a sum of many real
photons, which can pass through or reflect from semi-mirrors, and the
interference of the both light semi-wave is a result of the interaction of
all real photons, which build both semi-waves, with the standing waves of the
electromagnetic fields, created by stochastic exited virtual photons within
the closed contour between SmMr, FlMr1 and FlMr2. As this closed contour and
standing waves don't change at turning of the hole installation, than the
interference picture will not change because the interaction between
standing waves and real photons of both light semi-wave. In this physically
clear quantum explanation we understand that there is no need to use the
classical explanation and different mathematical transformatins because there
is a physically well-founded cause for an unchanged interference picture at
a turning of the hole installation of an angle $= 90^o$.

  The existence of the vacuum as some easy polarized ideal dielectric medium
is performance by the experimental observation of the Dopler's effect for
light as for sound and the limitation of the light velocity at its spreading
through the vacuum. When the closed contour is filled by some material
medium, having determined dielectric constant $\epsilon$, different from 1,
then the velocity u of the light within the medium, where $u = \frac{C}{n}$,
if $n = \sqrt{\mu.\epsilon}$ , where a $\mu$ is a magnetic permeability and
a $\epsilon$ is a electric permeability. Then by means of a relation (\ref{l4})
we could obtain :
\begin{equation}  \label{l13}
\quad u\prime = \frac{(\frac{C}{n} + v )}{ 1 + \frac{\frac{C}{n}.v}{C.C}} =
(\frac{C}{n} + v).(1 - \frac{1}{n}.\frac{v}{C}) \approx
\frac{C}{n} + w ,
\end{equation}

where w is a velocity of a carried light $ ( w = v.(1 - \frac{1}{n^2} =
v.(1 - \frac{1}{\varepsilon.\mu})$ When $\mu \approx 1$ then $ n^2 \approx
\varepsilon $ and therefore $ w \approx v.(1 - \frac{1}{\varepsilon})$

   If we see relation (\ref{i2}), then we understand that it is naturally
that the supplementary polarization of atoms and molecules within dielectric
medium delay the photon moving depending from velocity of moved medium. From
obtained above result about the light caring from a dielectric medium we
could understand that when there is some dielectric wave fluid thin a closed
contour, then its turning in the space will change the interference picture
of the interferometer.

  As all MicrPrts are excitements of the vacuum then every one of them would
can move freely through its ideal dielectric without any friction or
damping, that is to say without one to feel the existence of the vacuum.
Moreover, the existence of some MicrPrt in the vacuum distorts its ideal
crystalline lattice by its high density QntElcMgnFld, created by own FnSpr
ElmElcChrg. This natural distortion of the neutral FlcVcm molecular
close-packed lattice excites and ensures the gravitation field of the
ElmMicrPrt's mass, which by using same force show attention upon mass of
another ElmMicrPrt and upon its behavior. In such a naturally obvious and
physically clear way we understand why the force of the gravitation
interaction is determined by the self energy at a rest and mass.

\end{document}